\documentclass[10pt,conference]{IEEEtran}

\IEEEoverridecommandlockouts

\usepackage{cite}
\usepackage{amsmath,amssymb,amsfonts}
\usepackage{algorithmic}
\usepackage{graphicx}
\usepackage{textcomp}
\usepackage{xcolor}
\def\BibTeX{{\rm B\kern-.05em{\sc i\kern-.025em b}\kern-.08em
    T\kern-.1667em\lower.7ex\hbox{E}\kern-.125emX}}

\usepackage[T1]{fontenc}

\usepackage[utf8]{inputenc}
\usepackage[strict]{changepage}
\usepackage{framed}
\usepackage{pdfpages}
\usepackage[most]{tcolorbox}
\usepackage{tikz,lipsum,lmodern}
\usepackage{url}
\usepackage{tabularx}
\usepackage{booktabs}
\usepackage{adjustbox}
\usepackage{todonotes}
\usepackage{comment}
\usepackage[leftmargin=10pt,vskip=2pt]{quoting}
\usepackage{enumitem}

\begin{document}

\title{Leveraging Large Language Models for Cybersecurity Risk Assessment --- A Case from Forestry Cyber-Physical Systems}

\author{\IEEEauthorblockN{1\textsuperscript{st} Fikret Mert Gültekin}
\IEEEauthorblockA{
\textit{Chalmers University of Technology}\\
\textit{and University of Gothenburg}\\
Gothenburg, Sweden \\
fikretm@student.chalmers.se
}
\and
\IEEEauthorblockN{2\textsuperscript{nd} Oscar Lilja}
\IEEEauthorblockA{
\textit{Chalmers University of Technology}\\
\textit{and University of Gothenburg}\\
Gothenburg, Sweden \\
liljao@student.chalmers.se
}
\and
\IEEEauthorblockN{3\textsuperscript{rd} Ranim Khojah}
\IEEEauthorblockA{
\textit{Chalmers University of Technology}\\
\textit{and University of Gothenburg}\\
Gothenburg, Sweden \\
khojah@chalmers.se}
\and
\IEEEauthorblockN{4\textsuperscript{th} Rebekka Wohlrab}
\IEEEauthorblockA{
\textit{Chalmers University of Technology}\\
\textit{and University of Gothenburg}\\
\textit{| Carnegie Mellon University}\\
Gothenburg, Sweden | Pittsburgh, USA \\
wohlrab@chalmers.se}
\and
\IEEEauthorblockN{5\textsuperscript{th} Marvin Damschen}
\IEEEauthorblockA{
\textit{RISE Research Institutes of Sweden}\\
Borås, Sweden \\
marvin.damschen@ri.se}
\and
\IEEEauthorblockN{6\textsuperscript{th} Mazen Mohamad}
\IEEEauthorblockA{
\textit{RISE Research Institutes of Sweden}\\
\textit{| Chalmers University of Technology}\\
\textit{and University of Gothenburg}\\
Borås, Sweden | Gothenburg, Sweden\\
mazen.mohamad@ri.se}
}

\maketitle             

\begin{abstract}

In safety-critical software systems, cybersecurity activities become essential, with risk assessment being one of the most critical. In many software teams, cybersecurity experts are either entirely absent or represented by only a small number of specialists. As a result, the workload for these experts becomes high, and software engineers would need to conduct cybersecurity activities themselves. This creates a need for a tool to support cybersecurity experts and engineers in evaluating vulnerabilities and threats during the risk assessment process.
This paper explores the potential of leveraging locally hosted large language models (LLMs) with retrieval-augmented generation to support cybersecurity risk assessment in the forestry domain while complying with data protection and privacy requirements that limit external data sharing.
We performed a design science study involving 12 experts in interviews, interactive sessions, and a survey within a large-scale project.
The results demonstrate that LLMs can assist cybersecurity experts by generating initial risk assessments, identifying threats, and providing redundancy checks. The results also highlight the necessity for human oversight to ensure accuracy and compliance.
Despite trust concerns, experts were willing to utilize LLMs in specific evaluation and assistance roles, rather than solely relying on their generative capabilities.
This study provides insights that encourage the use of LLM-based agents to support the risk assessment process of cyber-physical systems in safety-critical domains.

\end{abstract}


\section{Introduction}

Cybersecurity is an increasingly important quality attribute in software-intensive systems, particularly in cyber-physical systems~\cite{ali2018cyber}.
Autonomous forestry is a particularly challenging domain, as it relies on advanced AI and networks of sensors to navigate complex terrain in remote locations.
Moreover, autonomous forestry systems often consist of multiple cyber-physical systems collaborating, which expands the attack surface and exposes communication protocols and control systems to potential malicious manipulation \cite{he2020challenges}.

Evolving regulations, including Regulation 2023/1230 on machinery \cite{eu-2023-1230} and the Cyber Resilience Act \cite{com-2022-454}, mandate rigorous safety and cybersecurity standards, demanding comprehensive risk assessments to mitigate cybersecurity threats and vulnerabilities.
Yet, organizations face a shortage of multidisciplinary experts in cybersecurity, AI, and regulatory compliance.
In this context, LLM-based tools have shown promise by summarizing regulatory requirements, generating draft reports to support expert-driven risk assessments \cite{collier2024risk} and other software artifacts~\cite{khojah2024beyond}.

While LLMs offer automation benefits, they require meticulous human oversight to mitigate data privacy risks, inaccuracies, and biases, and to ensure that essential expert review processes are not bypassed.

In this study, we explore whether locally hosted LLMs, such as Llama 2 \cite{touvron2023llama}, and retrieval-augmented generation (RAG) can support cybersecurity risk assessments. More specifically, we aim to understand to what extent and with what customizations an LLM can be used to support the risk assessment process, and examine how security and safety practitioners use the LLM, either to generate risk assessment documents in one shot or to support sub-activities that contribute to the risk assessment.
We applied a design science approach, involving 12 security and safety experts participating in interviews, interactive sessions with our LLM-based tool, and an evaluation survey, to iteratively refine our tool.
Our contributions are (i) an evaluation of an LLM-based tool with RAG to assist in cybersecurity risk assessments for mobile machinery in a large-scale context and (ii) an analysis of the benefits and challenges of integrating LLMs in risk assessments.
We show how companies can leverage LLMs in a lightweight fashion as supportive tools for cybersecurity risk assessment, and point the direction for developing future guidelines integrating LLMs in risk assessment processes.

\section{Background}

As part of a cybersecurity risk assessment in software engineering, the critical assets in a system are analyzed to identify security risks and derive security requirements.
Cybersecurity risk assessment involves (i) identifying and evaluating system vulnerabilities and threats, (ii) estimating their likelihood, and (iii) assessing potential impacts~\cite{62443}.
This involves comprehensive expert input and results in a prioritized list of risks guiding mitigation strategies.

Risk assessment documents typically include threat identification, asset protection, and mitigation strategies.
Standards developed by international organizations ensure that risk assessments are thorough and widely applicable.

\noindent
\subsection*{IEC 62443}
We relied on the IEC 62443 standard for cybersecurity \cite{62443}, which was originally developed for industrial automation and control systems. Its applicability across domains such as transportation~\cite{mirzai2024cybersecurity} demonstrates its versatility and makes it well-suited to address the complex and safety-critical security needs of autonomous forestry machinery.
IEC 62443 focuses on identifying potential threat agents and system vulnerabilities that could be exploited.
Techniques like \textit{attack trees} \cite{lallie2020review} can aid this analysis, as they systematically map out potential attack paths, highlighting vulnerabilities at each stage.
The framework also facilitates a structured \textit{Risk Determination} to assign \textit{Target Security Levels} that range from the most basic level of protection (SL 1) to the most sophisticated level of protection (SL 4).
Furthermore, the standard emphasizes the documentation of security requirements and mitigation strategies for each zone and conduit.

\noindent
\subsection*{The Project Context}
\label{sec:agrarsense}
This work is part of a project with 52 partners developing new agriculture and forestry technologies. 
It focuses on the cybersecurity of autonomous forestry machinery, among other use cases in the project.
The use case involves manually and autonomously operated machines on remotely planned worksites.
The autonomous forestry machinery domain comes with several regulatory challenges due to increasing automation and connectivity, which makes it an interesting area to explore LLM-assisted cybersecurity risk assessments in practice.

\section{Related Work}
AI has been used for various cybersecurity-related tasks, ranging from detecting security requirements \cite{mohamad2022identifying} to more complex tasks using LLMs, such as handling cyber threats and performing vulnerability analysis \cite{quinn2024applying}.
However, applications in cybersecurity require LLMs to have certain characteristics. For instance, Zhang et al. \cite{zhang2024llms} highlight that open-source LLMs such as Llama \cite{touvron2023llama} provide the adaptability to build on the current LLM architecture, which is valuable when performing cybersecurity tasks.

Beyond automating cybersecurity tasks, LLMs have also been used to assist security engineers in complex activities. Garza et al. \cite{garza2023assessing} proposed a system for answering questions related to threat behaviors and mitigation listed in the MITRE ATT\&CK framework, which can support security analysts and incident responders.
While other studies focus mainly on one activity, we cover multiple activities in the risk assessment process e.g., identifying threat actors, creating attack trees, and estimating threats likelihood.

In addition, cybersecurity risk assessment is underexplored in LLM research, but shares similarities with safety analysis. Researchers in the safety domain presented a framework using LLMs to automate the hazard analysis and risk assessment process by creating a pipeline and a sequence of prompts using prompt engineering techniques \cite{nouri2024welcome}. Kamile et al. have also explored a framework in risk assessment \cite{kamile2025riskassessment}. However, they emphasize that the main limitations of their framework come from the gap between academia and industry. Similarly, frameworks suggested by researchers remain theoretical without access to real-world attack data from industry \cite{kamile2025riskassessment}.
Therefore, one of the strengths of this work is bridging this gap and using real-world documents of risk assessments and standards used in forestry in the context of a large real-world project with 52 partners and 12 domain experts.

\section{Research Method}

We adopted a design science research approach~\cite{Hevner}.
This research method is particularly appropriate when solutions for a real-world problem need to be developed, even if the problem's context is not yet well understood.
For our goal, namely to investigate the potential use of LLMs for cybersecurity risk assessment, it was an applicable method, as it allowed us to iteratively study the problem's context and the feasibility of using LLMs.
Design science involves iteratively (i) understanding the problem, (ii) developing the design artifact, and (iii) evaluating the design artifact. In our study, the design artifact was a customized LLM specialized in using a RAG implementation for cybersecurity risk assessment in forestry automation.

\begin{figure}[ht]
  \centering  \includegraphics[width=\linewidth]{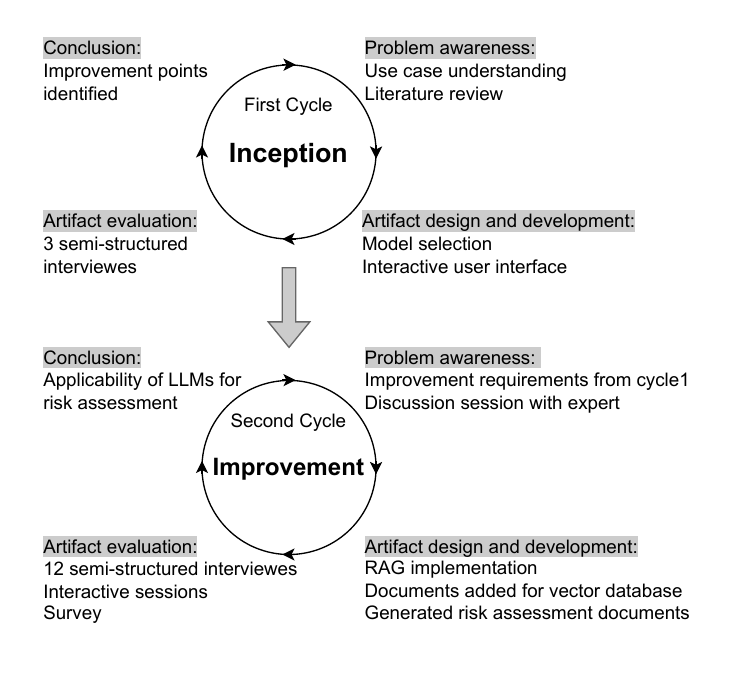}
  \vspace{-0.5cm}
    \caption{Our design science research approach}
    
  \label{fig:overview}

\end{figure}

The study involved 12 participants within the project. 
Details about the participants are shown in Tab.~\ref{tab:interview_data}.
We guaranteed anonymity, informed participants about the collection and use of data, and got full consent to carry out the study.

\begin{table}[ht]
\centering
\scriptsize
\caption{Participants' information, their participation in the cycles, and the interview duration in cycle 2.}
\begin{tabularx}{\linewidth}{p{0.08cm}p{2.79cm}p{0.97cm}p{0.88cm}p{0.66cm}p{1.25cm}} 
\toprule
ID  & Expertise & {Exp.} &  Company & Cycle & Interview duration \\
\midrule
1  & Cybersecurity & 4 years & A & 2 & 66 min \\ 
2  & Safety \& Cybersecurity & 11 years & B & 2 & 71 min  \\ 
3  & Cybersecurity & 2  years  & B & 1 \& 2 & 79 min \\ 
4  & Safety \& Cybersecurity & 30 years & B  & 1 \& 2  & 59 min \\ 
5 & Safety \& Cybersecurity & 23 years  & B  & 2 & 60 min  \\ 
6  & Safety & 18 years  & B & 2 & 39 min \\ 
7  & Cybersecurity & 13 years  & B & 2 & 51 min  \\ 
8  & Safety & 30 years  & B & 2 & 67 min  \\ 
9  & Safety & 20 years  & B &  1 \& 2 & 49 min \\
10  & Safety \& Cybersecurity & 20 years  & B &  2 & 48 min \\ 
11  & Safety \& Cybersecurity & 15 years  & B &  2  & 51 min\\ 
12  & Cybersecurity & 9  years  & B &  2 & 55 min\\ 
\hline
\end{tabularx}
\label{tab:interview_data}
\end{table}

An overview of our two-cycle research approach is depicted in Fig.~\ref{fig:overview} and described in the following.

\paragraph{Problem awareness}
In cycle 1, we reviewed related works to inform the artifact creation. We also had several discussions with the stakeholders of the project to analyze the use cases for cybersecurity risk assessment. We focused on understanding what LLM would best suit our purpose and what potential challenges we might face.

In the problem awareness phase of cycle 2, we discussed the problems that had arisen from cycle 1 and decided on how to improve them.
Particularly, we identified the need to focus on cybersecurity standards.
To obtain knowledge of IEC 62443, a discussion session with a cybersecurity expert was conducted. The expert directed the discussion and shared their knowledge of the standard by giving an example in the railway domain. The railway domain is also concerned with cyber-physical systems and, therefore, a suitable domain to get inspiration from for risk assessment in forestry.
The discussion session helped us to get insights into the need for a cybersecurity risk assessment approach that would be supported by a customized LLM.

\paragraph{Artifact development}
Llama 2 7B \cite{touvron2023llama} was selected as the main model. At the time of the study, it was the most up-to-date Llama version and a good choice for our case, as it could be deployed locally (to avoid sharing sensitive data with third parties), required moderate hardware requirements, had good performance, and was widely known by practitioners and researchers.
Note that bigger and better models are being continuously released, and the LLM landscape is changing quickly. Our goal was to investigate whether LLMs can be used for cybersecurity risk assessment in principle, which is why we focused on one of the state-of-the-art LLMs in this study.

In cycle 2, in the \textit{artifact improvement} stage, the LLM was customized by adapting a RAG architecture.
RAG was used to allow our LLM to fetch information from a curated database of closed-sourced data.
In comparison to fine-tuning, it is a less costly option and comes with minimal data processing requirements~\cite{lewis2020retrieval}.
RAG allows us to trace the contextual information used by the LLM to generate the outcome. 
It reduces hallucination and ensures that more domain-specific, focused responses are generated~\cite{gao2023retrieval}.
The RAG architecture employed a vector database with relevant documents and data.
In comparison to fine-tuning, the LLM got to focus on specific details and standards in the included documents without the need for extensive retraining.
Due to the proprietary nature of the documents and data, we cannot make them available.

For instance, we imported an existing risk assessment document following the IEC 62443 standard, which was an internal, project-specific document comprising a risk assessment from the railway domain.
It gave the LLM an example of a valid risk assessment.
Moreover, we included data about the project context (see Section~\ref{sec:agrarsense}), as well as data from the MITRE ATT\&CK techniques knowledge base. 
In total, 33 PDF documents were imported. 

Besides using RAG, we created prompt templates for the system and user prompts that aim to manage the overconfidence of the LLM and instruct it to ask for further verification if needed, as well as to provide a detailed explanation and more elaborate reasoning by following the automated chain-of-thought technique~\cite{cot}. This decision was made because our participants highlighted the need for a clear chain of evidence and elaboration on the reasoning of the generated output.

\paragraph{Artifact evaluation}
The artifact evaluation mainly focused on collecting qualitative data. We did not aim to fully automate the risk assessment, but always include humans in the loop, even when an LLM is used to generate at least a part of a risk assessment document. Therefore, qualitative feedback was essential to assess the LLM's usefulness and applicability to cybersecurity risk assessment.
Evaluation questions are in our supplementary material\footnote{\url{https://doi.org/10.5281/zenodo.16872695}}.

In both cycles, we conducted semi-structured interviews that consisted of open and closed questions. During the interviews, probing questions were asked to clarify unclear statements and to get more information~\cite{6449030}. 

All interview participants had previous experience with risk assessments (see Tab.~\ref{tab:interview_data}). The interviews lasted between 45 to 60 minutes and were conducted online or in person. Each interview was recorded and transcribed to facilitate the analysis.

In cycle 2, we performed interactive sessions to observe the interviewees' interactions with the tool. The interviewees were asked to perform a risk assessment of our project (See Section \ref{sec:agrarsense}) while interacting with our model via prompts that they find important to assist them in the risk assessment process. To give them a head start, we provided an optional first prompt to retrieve the primary assets of the project.

We observed their prompting style and reactions to the tool.
The participants spent 10--15 minutes trying the model. The interviewees were instructed to explore the LLM with different tasks they would need support with when conducting a risk assessment. While they were entering prompts and reviewing the outputs, all inputs and outputs were recorded for later analysis. 

After the interactive sessions, we collected data and feedback on our model via a survey that mainly consisted of Likert-scale questions. 
The questions focused on the usefulness, completeness, reliability, level of detail, and relevance of the generated output. 
We also included demographic questions. 

\paragraph{Derivation of findings and conclusion}
Using content analysis, the transcripts and recordings were thoroughly analyzed, and each statement was assigned a code. 
Pre-defined codes were created for the evaluation criteria, e.g., the level of detail. During the coding process, additional codes were created in an emerging fashion, resulting in four themes that we present in detail in Section \ref{sec:results}.

\section{Results}
\label{sec:results}

We present the requirements for LLMs used in risk assessment and the evaluation of our customized model. 

\subsection{Requirements for LLM-supported risk assessment}
\label{sec:res:llmReqs}

In the first cycle, using three interviews with cybersecurity and safety experts, we elicited \emph{activities} that LLMs are expected to support and the \emph{characteristics} that they should have in order to be helpful.

The most important activity that can assist practitioners during the risk assessment process was \textbf{identifying security threats and risks}. More specifically, the LLM should enable the detection of potential threats that can occur within a specific context. Identifying threats is particularly challenging in forestry due to its remote and complex nature, where security threats and risks are not always immediately obvious, especially given the influence of unpredictable natural factors \cite{mohamad2024cybersecurity}.

There was also an emphasis on the importance of a \textbf{dynamic interaction with the LLM} to {enhance the risk management process}. This includes supporting key activities such as \textbf{identifying threat actors and assessing potential risks} in terms of severity and likelihood.

Finally, one challenge that practitioners often face with risk assessment, especially when using automated tools, is the completeness of the assessment.
Therefore, the participants stressed that a useful use case that LLMs could help with is \textbf{evaluating the completeness of the overall process}. This could suggest that the practitioner considers overlooked scenarios and reflects on the coverage of the risk assessment.
The participants indicated that this use of LLMs, along with the aforementioned activities, can serve as a redundancy check by examining whether the risk assessment conducted by the practitioner covers the risks identified and assessed by the LLM, or whether any crucial elements were missed.

Beyond LLM capabilities, some characteristics of LLMs are crucial when leveraged for cybersecurity risk assessment. Since the risk assessment process can be specific to the domain, the LLM should be \textbf{context-aware}. The context encompasses the systems and machines involved, the environment, and unique scenarios that are linked to the domain. For instance, shuttles are exposed to certain risks when idle in the woods. 
The participants also emphasized the importance of following functional \textbf{safety and cybersecurity standards} that were related to the project. The goal of following a standard is to ensure that the outcome is generated using standardized methods and processes ensuring its \textbf{consistency} regardless of who conducts the risk assessment. In the project, the standard that was the most suitable among the investigated standards was IEC 62443 \cite{62443}. While IEC 62443 is a cybersecurity standard, it has indirect safety implications since it covers cybersecurity threats that cause hazardous events.

However, all participants were familiar with LLMs' tendency to hallucinate and be over-confident, which can be problematic in a safety-critical domain such as forestry. So, they acknowledged that while ideally the LLM's responses are expected to be complete and fully capture all possible risks without human oversight, they did not trust it to reach this level just yet.
However, to improve the trust and the reliability of the outcome, the participants noted that it is crucial for an LLM to be transparent and \textbf{request verification} from a security expert when needed. To enable human verification, the LLM is expected to explain the outcome (e.g., identified risks) along with an elaboration of the \textbf{reasoning behind its outcome} and its conclusions. This way, the practitioner will feel more comfortable using the LLM as an automation tool, and trust that it will delegate crucial tasks that require human intervention to the experts.

\subsection{Evaluation of using an LLM for risk assessment} \label{sec:eval}
To apply the requirements we elicited in the previous section, we customized the Llama model by using a RAG architecture with a vector database. 
We conducted 12 interviews with safety and cybersecurity experts to i) evaluate the usability and usefulness of the customized Llama model for risk assessment in general, ii) evaluate the quality of the generated risk assessment document by the Llama model, and iii) suggest further improvements to the Llama model accordingly.

\subsubsection{Evaluation of the Llama 2 model}
After adapting the Llama 2 model using a RAG architecture and prompt engineering, we performed interactive sessions with our model to assess the model's \textit{usability} and \textit{usefulness} for various activities.
All of the interviewees participated in the sessions and were asked to prompt our model to perform different tasks they would need support with in a risk assessment process. We suggested that participants use the LLM to identify primary assets, as a way to initialize the interaction. However, we left a lot of room for flexibility, so that they could freely decide what prompts to create and what activities to focus on.

\begin{table}[!ht]
\scriptsize
    \caption{Characteristics of prompts provided by the participants during the interaction session with our model.}
    \begin{tabularx}{\linewidth}{p{1.5cm}p{1.5cm}p{1.5cm}p{2.75cm}}
    \toprule

    \textbf{Participant} & \#\textbf{Prompts} & \#\textbf{Updates} & \textbf{Avg. prompt length} \\ 
    \midrule
    P1 & 3 & 2 & 293.30 \\ 
    P2 & 11 & 4 &  141.09 \\ 
    P3 & 5 & 0 &  122.20 \\ 
    P4 & 3 & 0 & 173.00 \\ 
    P5 & 9 & 2 &  58.30 \\ 
    P6 & 5 & 3 &  95.80 \\ 
    P7 & 4 & 1 &  218.25 \\ 
    P8 & 5 & 3 & 97.40 \\ 
    P9 & 2 & 0 & 78.00 \\ 
    P10 & 3 & 1 &  50.30 \\ 
    P11 & 5 & 3 &  101.40 \\ 
    P12 & 10 & 4 &  93.40 \\ 
    \midrule
    \textbf{Total} & 65 & 23 & 1522.44 \\ 
    \textbf{Average} & 5.42 & 1.92 &  126.87 \\ 
    \textbf{Max} & 11 & 4 & 293.30 \\ 
    \textbf{Min} & 2 & 0 & 50.30 \\ 
    \bottomrule
\end{tabularx}
\label{tab:demo_data}
\end{table}

In Table~\ref{tab:demo_data}, we provide data on the prompts that were written by our interview participants.
The participants queried 65 prompts in total, of which 23 were updated by rephrasing the prompt or adding a more detailed description of the query in an attempt to refine or improve the output of the model. The interactions were generally short, with an average of 5 prompts per participant.

In addition to identifying primary assets, participants decided to prompt the model for other use cases (see Table~\ref{tab:demo-usage}). The most common use cases were the identification of primary assets and identifying threats, followed by understanding the system and general information retrieval to obtain knowledge about the context, domain, and security and safety concepts in general.

\begin{table*}[!ht]
\scriptsize
    \caption{Characteristics of prompts provided by the participants during the interaction session with our model.}
    \begin{tabularx}{\linewidth}{p{3.1cm}p{0.5cm}p{1.9cm}X}
    
    \toprule
    \textbf{Activity} & \textbf{Freq.} & \textbf{Area} & \textbf{Example prompt} \\
    \midrule

    Identify primary assets & 7 & Security & What are potential primary assets to consider when doing a risk assessment of [system]?  \\
    Identify threats &  7 & Security  & Which MITRE ATT\&CK techniques relate to spoofing when the primary asset is a drone?  \\
    Understand the system &  5 & Security, Safety  &  List the core functions of the [project] autonomous shuttle.  \\
    Information retrieval & 4  &  Security, Safety & What are the primary security attributes? \\
    Identify risks &  2 & Safety  & How do the environmental factors influence the risk of safety at the forestry site [...]? \\
    Create attack trees & 2  & Security  &  Description the attack tree for a ransomware attack targeting the autonomous shuttle \\
    Standards and practices &  2 &  Security, Safety &  Please provide information regarding relevant EU-directives for [project]. \\ 
    Guide security assurance &  2 & Security  &  I have an encryption key for a TLS certification. How should I protect it? \\
    Identify threat actors & 1  & Security  & List threat agents relevant to survey drones.  \\
    Understand risks and threats impact &  1 &  Security, Safety & Given [context], give me some types of image quality degradation due to the noise jamming attacks on the wireless communication? \\ 
    \bottomrule
\end{tabularx}
\label{tab:demo-usage}
\end{table*}

\subsubsection{Evaluation of the LLM-generated risk assessment}
\label{sec:eval-doc}
We evaluated the model's ability to automatically generate a complete risk assessment document. The following findings cover the results of the content analysis that we performed on the 12 interviews of the second cycle, as well as the Likert-scale results from the survey the participants filled out after the interview.

The analysis resulted in themes that focus on the quality aspects of the LLM-generated risk assessment document that were discussed by the interviewees. 

\begin{figure}[ht]
    \centering
    \includegraphics[width=\linewidth]{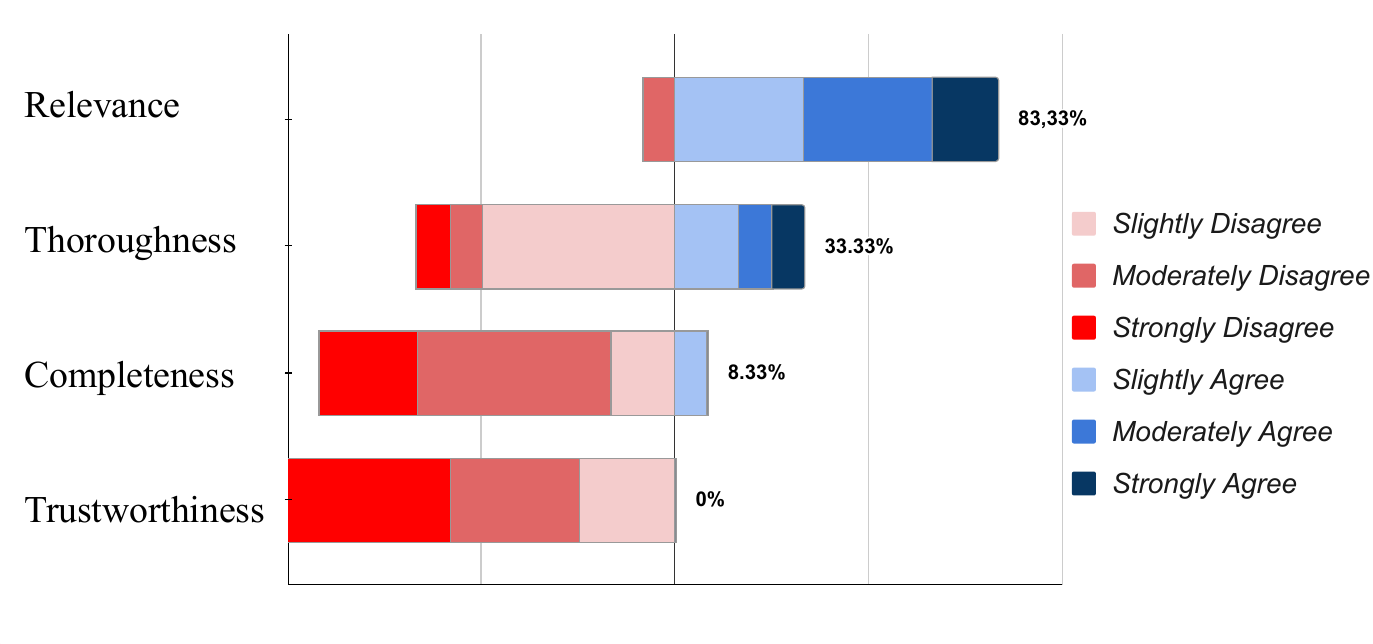}
    \caption{Perceived fulfillment of selected qualities in generated report.}
     \label{fig:survey-results}
\end{figure}

The main quality aspects that the interviewees focused on were the \textit{relevance} (to the specific domain and activities), \textit{thoroughness} (the extent of details), \textit{completeness} (the extent to which output contained all necessary information), and \textit{trustworthiness} (the extent to which participants trust the output) of the risk assessment document. The statements of the participants aligned with the survey results, which are depicted in Fig.~\ref{fig:survey-results}.

\paragraph{Trustworthiness} The participants' main concern was with the trustworthiness of some parts of the risk assessment.
Participants expressed their concerns regarding overreliance on the generated risk assessment. Six participants said that they would not trust the generated risk assessment given that it included inaccuracies, inconsistencies, and missing information. This also showed in the survey where none of the participants thought that the generated risk assessment was trustworthy (Fig.~\ref{fig:survey-results}).

\begin{quoting}
\small
    \textit{``If you give this to the hands of someone who is not at all a cybersecurity expert and blindly trusts what it gives, then it’s dangerous.''}
    \textit{--- P4}
\end{quoting}

The perceived level of trustworthiness is connected to eight participants thinking that parts of the document were inaccurate. The document lacked adherence to the definitions and scope of different risk assessment elements. For instance, primary assets should be assigned to the core functions of the system according to IEC 62443, whereas in other standards it can be mapped to physical components or even use cases. In the primary assets generated, there was a mix of different functions, use cases, and physical components of the system, making the primary and secondary assets inconsistent and inaccurate.
Moreover, when it came to calculations of numeric measures such as the damage potential and risk level, the numbers seemed inaccurate and arbitrary, and the rationale behind them would often be missing or not make sense to the participants.

\paragraph{Completeness} Seven participants thought that the threat actors, the description of the primary assets, and possible attack scenarios in the attack tree were correct and can be seen in an actual risk assessment for the project of our case company. They also appreciated the document followed the structure of the IEC~62443 \cite{IEC61508} standard, and included an overview of the risk assessment, threat actors that are accurate for the project, and reasoning behind the identified threats.
 
\begin{quoting}
\small
\textit{``I felt that the structure was quite good. It has the main ingredients [of the risk assessment]''}
\textit{--- P7}
\end{quoting}

Nine participants pointed out missing information that they expect to see in a risk assessment document. More specifically, a system description, a diagram that illustrates the system, and information about the specific hardware that the software is executed on were missing. The participants also highlighted that while the document followed the standard structure, it lacked several steps in the content of the assessment. For instance, an asset owner is needed for each identified asset.

\paragraph{Thoroughness and relevance} five participants found some parts of the generated risk assessment document to be too generic and applicable to other domains and systems, not necessarily the project, making them hard to understand. For example, the attack tree was generic to unauthorized users rather than a threat actor specific to the system. 
The identified assets were related to general components of any system (e.g., Software Component or Hardware Vehicle) and did not consider the core functions of the specific system, in which a drone is used with functionalities such as navigation, AI perception, and automated decision-making, among others.

Similarly, identified threats and impacts were mapped to a single physical component of the system (e.g., threat X impacts the drone, threat Y impacts the shuttle), while components in cyber-physical systems are connected and can therefore be collectively impacted  by the same threats.

Although the risk assessment was generic, six participants acknowledged that the overall risk assessment was still useful and relevant to apply in the forestry domain and the case project. In fact, some have also emphasized that the generic nature of the identified risks can be an advantage since otherwise, practitioners tend to focus on too specific risks and miss the obvious and more general ones. This indicates that the output of our model can be a good starting point that needs to be reviewed and modified to be more specific. 

\begin{quoting}
\small
\textit{``That’s interesting because [the primary assets] are relevant, but very generic, so it is they are suitable for almost any system, but that’s not a bad thing because you may miss the obvious things as well.''}
\textit{--- P4}
\end{quoting}

\subsubsection{Suggestions for further improvements}
Several participants provided suggestions to improve the generated output and
new use cases or standards:

\begin{quoting}
\small
\textit{``I think it would be interesting to apply it to machinery directive [and] ISO 12100.''}
\textit{--- P8}
\end{quoting}

The participants also mentioned that the risk assessment would be more trustworthy if the following is provided: an explanation of the perceived definition of each element (e.g., secondary assets), and the reasoning behind different calculations (e.g., risk level). The participants believe that this can be fixed with prompt engineering where some participants observed during the interactive session that the longer time they spend on constructing their prompt, the better the outcome became.

Other suggestions concerned the design of the LLM. A conversation-level context awareness was requested to avoid the need to bulk-prompt the LLM to do a complex task such as generating a whole risk assessment document, which requires a large context. Another suggestion was to add more relevant documents to the database that our RAG architecture uses such as safety standards and regulations, as well as security assurance documents to allow the LLM to suggest threat mitigation strategies.

\subsubsection{Usefulness of the LLM with various activities}

In the interviews and interactive sessions, our participants indicated that the LLM can be useful for cybersecurity risk assessment.
In particular, nine participants found it beneficial to have an LLM assess and provide suggestions to improve an existing risk assessment. Such an LLM-driven assessment focused on criticizing specific parts of the risk assessment such as asset protection and threat mitigation.
Further, the LLM could identify gaps in the protection and mitigation strategies developed by practitioners. The participants also mentioned that using it for such purposes can serve as an initial sanity check and a verification of their risk assessment, which can increase their confidence.

Other activities that were highlighted are the ability of the LLM to perform monotonous and time-consuming tasks, such as data analysis and information retrieval. 
Another aspect of usefulness appreciated by four participants was the potential to have the LLM act as a college or sparring partner when performing highly creative tasks like brainstorming or threat identifications.

Collectively, this demonstrates the ability of LLMs to act as a supplementary tool to established practices for risk assessment. 83.33\% of the participants indicated that they would use our LLM to support them in their line of work.
However, some key challenges would need to be combated to further increase the efficiency and usability. As with current LLMs, reliability and correctness are still an issue, especially in the more niche domains~\cite{wang2023assessingreliabilitylargelanguage}. Without improving the ability to produce correct information and provide coherent reasoning, practitioners simply cannot rely on the output with enough certainty. This was pointed out by three participants who mentioned that with the current possibility of inaccurate outcomes, extra manual work would be needed to check the reliability of the outcome.

\section{Discussion}

In this section, we point to gaps in LLM usage for risk assessment, and describe how an agentic system can be applied to improve its support in the risk assessment process. We also discuss the threats to validity.

\subsection{Suitability of LLMs for risk assessment}
Creating a risk assessment document requires in-depth knowledge of the system, its environment, and its security requirements to ensure that the assessment is relevant and accurate. Therefore, the LLM needs a lot of context in order to make the assessment relevant (see Section~\ref{sec:eval-doc}). However, our results indicate that even with the RAG implementation, the LLM's output lacked completeness and specificity. One factor was that some documents we included in our vector database were indeed applicable to various cyber-physical systems (e.g., the IEC 62443 standard) but not specific to forestry due to the lack of cybersecurity standards in this domain. 

Moreover, cybersecurity in safety-critical systems is a sensitive area where the potential consequences of errors are significant. We observed a generally low trust among participants (see Figure~\ref{fig:survey-results}). We highlight the need for transparency and trustworthiness before these technologies can be widely adopted in high-risk sectors. A key aspect of building this trust is ensuring that the source of any generated content, whether from an LLM, human input, or a combination of both, can be traced \cite{tracability2024trust}. This is important to identify the origin of a security insight or recommendation and to evaluate the reliability of the output. For LLMs to be trusted in cybersecurity for safety-critical systems, developing systems for source verification and transparency will be crucial~\cite{trust2025cybersecurity}.

\subsection{Moving towards a standards-based agentic system}
We argue that agentic systems can address the gaps we found in using an LLM-based system by (i) decomposing the risk assessment task into specialized roles, (ii) providing transparency and traceability through structured workflows and verifiable outputs, and (iii) enforcing human-in-the-loop checkpoints.

An agentic system for cybersecurity risk assessment can benefit from using multiple agents. Each agent is responsible for a specific risk assessment activity in accordance with a standard (IEC 62443 in our case). For example, one agent could focus on identifying primary assets and building structured cybersecurity artifacts (e.g., data flow diagrams or software bills of materials), while another specializes in generating and refining potential threats by utilizing these artifacts. This aligns with our participants' emphasis on adhering to established standards and supports their need for a structured process that can improve trust in the assessment outcomes.

Moreover, we found that providing context through RAG could result in retrieving general information that did not add value to the generated output (see Section~\ref{sec:eval-doc}). Therefore, to improve the context and information used by the different agents that use RAG (e.g., to retrieve relevant threat patterns from MITRE ATT\&CK), we suggest that a preprocessing agent would ensure that RAG results in providing relevant contextual information by removing ambiguity and irrelevant data. 
Then, to improve traceability, each agent can log the document chunks used to form the output, including references to standards, system specifications and requirements, or prior assessments. This allows cybersecurity experts to assess the reliability of the generated suggestions.
In addition, one agent can be responsible for collecting different evidence related to the risk assessment and treatment, and integrating them into an assurance framework with the help of the traceability output. This helps cybersecurity experts in managing evidence in their organizations, which is a challenge in current practice~\cite{mohamad2024evidence}.

Our participants emphasized the importance of integrating human-in-the-loop mechanisms for improving their trust and the LLM performance. 
In an agentic system, an orchestrator agent can be used to enforce a workflow that is compatible with the standard in use and trigger human-in-the-loop checkpoints. The cybersecurity expert acts as an auditor of each agent's output and ensures that legal and ethical requirements are met. For example, when an agent specifies a list of risks, the human can verify this list before using it by the next agent that would calculate the risk levels. 
As future work, we plan to develop an agentic system based on our findings, as our results indicate that specialized agents have the potential to mitigate limitations by using only LLMs and support cybersecurity experts in the risk assessment process.

\subsection{Threats to Validity}

To improve \textbf{internal validity}, we selected
experts with an average of 15 years of experience working with risk assessment, which ensured that our insights were reliable and relevant to our research.
Moreover, to reduce biases and misunderstandings when interpreting the data, we used data triangulation through interactive sessions, semi-structured interviews, and a survey. 
We also consider the LLM's non-deterministic nature by running each prompt five times and selecting the most representative response.

In terms of \textbf{external validity}, we focus only on forestry domain. However, the procedure outlined in the study could easily be replicated in other domains, as the training data primarily came from outside the forestry domain. 
One threat was the lack of diversity and the sample size of the interview participants. We recommend that future studies involve different domains and more participants to enhance the generalization of the results.
Lastly, while the choice of model can impact our findings, we mainly focused on the interaction and the assistance of LLMs. 

\textbf{Construct validity} was ensured by clear definitions of regularly occurring concepts and codes. This ensured a common understanding among researchers, interview participants, and domain experts.

\section{Conclusion and Future Work}
This design-science study has demonstrated the potential of using LLMs to support cybersecurity risk assessments in autonomous forestry machinery.
The prototyped LLM-based tool, integrating RAG, has shown promising results as a supportive assistant for practitioners.

The involved experts found the tool useful for conducting monotonous tasks, generating initial cybersecurity risk assessments, and criticizing manually created risk assessments.
We found that LLMs can identify aspects overlooked by humans and perform sanity checks, which makes it feasible to collaborate with LLMs to support in the cybersecurity risk assessment process of a real-world, large-scale project. However, certain challenges limit the usefulness of LLMs when used for automating the risk assessment process without human oversight.
Future work will build on our findings and conduct another design science cycle in which the artifact is a tool that utilizes a multi-agent architecture.
We also aim to explore the applicability in other safety-~and security-critical sectors.
We see LLM-based tools as a promising approach to alleviating the shortage of experts needed to perform risk assessments for compliance with upcoming regulations.

\section*{Acknowledgements} 
We thank Komatsu Forest AB's valuable insights as the AGRARSENSE coordinator and the forestry use case leader.
AGRARSENSE is supported by the Chips JU and its members, including top-up funding from Sweden, Czechia,
Finland, Ireland, Italy, Latvia, Netherlands, Norway, Poland and
Spain (Grant Agreement No. 101095835).
The views expressed in this document are the authors' sole responsibility and do not necessarily reflect the views or position of the European Commission.
Neither the authors, the AGRARSENSE consortium, nor the Chips JU are responsible for the use that might be made of the information contained in this document.

This work was also partially supported by the Wallenberg AI, Autonomous Systems and Software Program (WASP) funded by the Knut and Alice Wallenberg Foundation.


\bibliographystyle{ieeetr}
\bibliography{bibliography}
\end{document}